\documentclass[a4paper,aps,groupedaddress,showpacs,showkeys,nofootinbib]{revtex4}
\RequirePackage[colorlinks,hyperindex]{hyperref}
\RequirePackage[english]{babel}
\RequirePackage[latin1]{inputenc}
\RequirePackage[T1]{fontenc}
\RequirePackage{mathrsfs}
\RequirePackage{amsmath}
\RequirePackage{amssymb}
\RequirePackage{amsbsy}
\RequirePackage{color}
\RequirePackage{bm}

\newcommand{\be}{\begin{equation}}
\newcommand{\ee}{\end{equation}}
\newcommand{\bea}{\begin{eqnarray}}
\newcommand{\eea}{\end{eqnarray}}

\def\de#1/de#2{\frac{\partial {#1}}{\partial {#2}}}
\hypersetup{colorlinks=true,breaklinks=true,urlcolor=blue,linkcolor=red}
\begin{document}
\title{Axially-Symmetric Exact Solutions for Flagpole Fermions with Gravity}
\author{Roberto Cianci\footnote{E-mail: cianci@dime.unige.it},
Luca Fabbri\footnote{E-mail: fabbri@dime.unige.it},
Stefano Vignolo\footnote{E-mail: vignolo@dime.unige.it}}
\affiliation{DIME Sez. Metodi e Modelli Matematici, Universit\`{a} di Genova\\
via all'Opera Pia 15 - 16145 Genova, ITALY}
\date{\today}
\begin{abstract}
The Lounesto classification splits spinors in six classes: I, II, III are those for which at least one among scalar and pseudo-scalar bi-linear spinor quantities is non-zero, its spinors are called regular, and among them we find the usual Dirac spinor. IV, V, VI are those for which the scalar and pseudo-scalar bi-linear spinor quantities are identically zero, its spinors are called singular, and they are split in further sub-classes: IV has no further restrictions, its spinors are called flag-dipole; V is the one for which the spin axial-vector vanishes, its spinors are called flagpole, and among them we find the Majorana spinor; VI is the one for which the momentum antisymmetric-tensor vanishes, its spinors are called dipole, and among them we find the Weyl spinor. In the quest for exact solutions of fully-coupled systems of spinor fields in their own gravity, we have already given examples in the case of Dirac fields \cite{CFV2}, and Weyl fields \cite{CFV1} but never in the case of Majorana or more generally flagpole spinor fields. Flagpole spinor fields in interaction with their own gravitational field, in the case of axial symmetry, will be considered. Exact solutions of the field equations will be given.
\end{abstract}
\maketitle
\section{Introduction}
In contemporary physics, the quest for exact solutions is an endeavour that is scarcely pursued. In spite of the rarity of exact solutions, we have managed to handle tools powerful enough to extract a great deal of information out of physical systems. But although such information may be sufficient to make central predictions, nothing is as informative as exact solutions themselves. To be precise, some exact solution has actually been found in the past.

One example is the case of the plane-wave solutions of Klein-Gordon or Dirac, or even Maxwell equations. However, single, isolated fields do not really exist in nature, since every field has an energy and therefore every field should in principle have at least the interaction with its own gravitational field. Allowing gravity into the picture, another example arises, which is given by the Kerr solution, encompassing the gravitational and electrodynamic fields of a massive charged material distribution. But just the same, this solution is exact in gravity and electrodynamics, while the matter distribution is concealed within the central singularity, and so it is not a complete solution either.

Of course studying a fully-coupled system of field equations, that is one for which the gravitational field is sourced by an energy density given by a field whose field equations contains the connection of the gravitational field, one need to be equipped with a large amount of patience, for solutions are hard to get. Some simplification must therefore be implemented. The simplest system we could study is that of neutral matter distributions, so to have gravitation alone. When the matter distribution is of spinorial type, some solution has been found, albeit in special cases.

To better appreciate the way in which these special cases are arranged, it is necessary to recall that spinor fields can in general be classified in terms of various categories. One such instance is that of Wigner, which examines matter in terms of the two Casimir operators of the full 
Poincar\'{e} group, that is the momentum vector and Pauli-Lubanski axial-vector, classifying particles according to whether they are massive or massless and in means of their quantized spin states. This classification is still widely used today, but it has recently been accompanied, and refined, by a number of other classifications, which employ specific mathematical tools given by the tensor bi-linear spinor quantities, obtained by bracketing between conjugate spinors the elements of the Clifford algebra \cite{L,HoffdaSilva:2017waf,daSilva:2012wp,Cavalcanti:2014wia,Ablamowicz:2014rpa,Fabbri:2016msm}. Different papers among \cite{L,HoffdaSilva:2017waf,daSilva:2012wp,Cavalcanti:2014wia,Ablamowicz:2014rpa,Fabbri:2016msm} contain different details of the possible classifications, and it is not the place here to enter in these technicalities.

However, all papers \cite{L,HoffdaSilva:2017waf,daSilva:2012wp,Cavalcanti:2014wia,Ablamowicz:2014rpa,Fabbri:2016msm} have generally the same classification scheme, and according to such a classification scheme, spinors can be split into two large classes, one of which given when both scalar bispinor quantities are zero, and thus called class of \emph{singular} spinors, and the other given when either one of the two scalar bispinor quantities is non-zero, and thus called class of \emph{regular} spinors. Additionally, each class can be further split in three sub-classes, so that regular spinors are split in type-I and type-II, type-III while singular spinors are split into type-IV, type-V and type-VI. Type-I, II and III are altogether called Dirac spinors (differences between classes are based on details that once again are irrelevant in the present context). Type-IV are called \emph{flag-dipole} \cite{daRocha:2013qhu,VFC} with type-V being the flagpole (those for which the spin axial-vector vanishes, and containing as a particular case the self-conjugated Majorana spinors) and type-VI being the dipole (those for which the momentum antisymmetric-tensor vanishes, and containing as a particular case the single-handed Weyl spinors).

With this classification in sight, the fully-coupled exact solutions found so far can be categorized. So to begin, some fully-coupled exact solutions for the case of Dirac fields have been found in \cite{CFV2}. In \cite{CFV1} we discussed the fully-coupled exact solutions of the Weyl class. Because the Weyl case is simpler than the Dirac case, we also found all of them.

There is yet no fully-coupled exact solution corresponding to spinors belonging to the flagpole class. Looking for solutions that could exhaust all classes, flagpole class should be studied as well. Searching for fully-coupled exact solution of flagpole type is what we will do in the present paper.

There paper is organized with a first section in which we will set the notation and convention for the basic definitions, and where we will give the field equations we will employ. Section III will be about writing everything in the axially-symmetric configuration. In Section IV we will proceed in solving the resulting field equations. Section V will contain a final summary.
\section{Notations and field equations}
In the paper, partial derivatives of a given function $f(x^h)$ are indicated by $f_{x^h}:=\frac{\partial f}{\partial x^h}$. Latin and Greek indices run from $1$ to $4$. The metric tensor of the space-time is denoted by $g_{ij}$ while the tetrad field associated with a given metric is indicated by $e^\mu_i$ in such a way that $\eta^{\mu\nu}=g_{ij}e^i_\mu e^j_\nu$ with $\eta^{\mu\nu}=\mathrm{diag}(-1,-1,-1,1)$ being the Minkowski metric, and with inverse $e^i_\mu$ verifying therefore $e^i_\mu e^\mu_j=\delta^i_j$ and $e^\mu_i e^i_\nu= \delta^\mu_\nu$ in terms of Kronecker delta. Dirac matrices are indicated by $\gamma^\mu$ and $\Gamma^i := e^i_\mu\gamma^\mu$ with $\gamma^5 =i\gamma^4\gamma^1\gamma^2\gamma^3$ and chiral representation is used. The spinorial-covariant derivatives of a Dirac field $\psi$ are expressed as 
\begin{equation}\label{defdsm}
D_i\psi=\psi_{x^i}-\Omega_{{i}}\psi
\end{equation}
where the spinorial-connection coefficients $\Omega_i$ are given by
\begin{equation}\label{1.1}
\Omega_i := - \frac{1}{4}g_{jh}\omega_{i\;\;\;\nu}^{\;\;\mu}e_\mu^j e^\nu_k\Gamma^h\Gamma^k
\end{equation}
with $\omega_{i\;\;\;\nu}^{\;\;\mu}$ being the coefficients of the spin-connection associated through the relation
\begin{equation}\label{1.2}
\Gamma_{ij}^{\;\;\;h} = \omega_{i\;\;\;\nu}^{\;\;\mu}e_\mu^h e^\nu_j + e^{h}_{\mu}\partial_{i}e^{\mu}_{j}
\end{equation}
to the linear connection $\Gamma_{ik}^{\;\;\;j}$ as usual. 

We will consider a Dirac field coupled to gravity in the Einstein general relativity theory. The Dirac Lagrangian is
\begin{equation}\label{1.1bis}
L_\mathrm{D} =\frac{i}{2}\left( \overline{\psi}\Gamma^iD_i\psi-D_i\overline{\psi}\Gamma^i\psi\right)-m\overline{\psi}\psi
\end{equation}
where $m$ is the mass of the spinor. Field equations are obtained by varying the Einstein--Hilbert plus the Dirac Lagrangian with respect to the tetrads and spinor fields: they turn out to be the Einstein equations
\begin{equation}\label{1.3a}
R_{ij} -\frac{1}{2}Rg_{ij}= \frac{i}{4} \left( \overline{\psi}\Gamma_{(i}{D}_{j)}\psi - {D}_{(j}\overline{\psi}\Gamma_{i)}\psi \right)
\end{equation}
and the Dirac equations
\begin{equation}\label{1.3b}
i\Gamma^{h}D_{h}\psi - m\psi=0
\end{equation}
where $R_{ij}$ and $R$ denote the Ricci tensor and the curvature scalar associated with the Levi--Civita connection. Notice that by means of the Dirac equations, the Einstein equations \eqref{1.3a} can be written in the equivalent form
\begin{equation}\label{1.3c}
R_{ij} = \frac{i}{4} \left( \overline{\psi}\Gamma_{(i}{D}_{j)}\psi - {D}_{(j}\overline{\psi}\Gamma_{i)}\psi \right) - \frac{m}{4}\overline{\psi}\psi\/g_{ij}
\end{equation}
which is easier to treat.

It is important to remark that no torsion has been considered, but this amounts to no loss of generality because the spinors we are going to consider are flagpole spinors, those having the axial-vector bispinor quantity, that is the Pauli-Lubanski axial-vector, identically equal to zero. Because torsion couples to the dual of the axial-vector bispinor quantity, which is then also zero, then torsion is zero on-shell.

We are now ready to write all field equations in {\it cylindrical coordinates}.
\section{Spinor Fields in Lewis-Papapetrou space-time}
We assume a Lewis-Papapetrou metric in \emph{cylindrical coordinates} of the form 
\begin{equation}\label{Lewis}
{ds}^{2}= -B^2 (d\rho^2 +dz^2) -\rho\/P (-W\,dt + d\varphi)^2 + \frac{\rho}{P}\,dt^2
\end{equation} 
where the functions $B(\rho,z)$, $P(\rho,z)$ and $W(\rho,z)$ depend on $\rho$ and $z$ only. The tetrad field associated to \eqref{Lewis} is
\begin{equation}\label{defcoframe}
e^1 = B\,d\rho, \quad e^2= B\,dz, \quad e^3 = \sqrt{\rho\/P}\left(-W\,dt +d\varphi\right), e^4 = \sqrt{\frac{\rho}{P}}\,dt
\end{equation}
and its dual is
\begin{equation}\label{defframe}
e_1 = \frac{1}{B}\,\frac{\partial}{\partial \rho}, \quad 
e_2 = \frac{1}{B}\,\frac{\partial}{\partial z}, \quad 
e_3 = \frac{1}{\sqrt{\rho\/P}}\,\frac{\partial}{\partial \varphi}, \quad 
e_4 = \sqrt{\frac{P}{\rho}}W\,\frac{\partial}{\partial \varphi} + \sqrt{\frac{P}{\rho}}\,\frac{\partial}{\partial t}
\end{equation}
as it is easy to check.

In this paper, we intend to impose these symmetries also on the bilinear quantities arising from the spinor field: to this purpose, we insist on the vanishing of the Lie derivative for both the velocity vector $V=\bar\psi\gamma^\mu\psi$ and the Pauli-Lubanski axial-vector $S=\bar\psi\gamma^\mu\gamma^5\psi$, namely $L_{\frac{\partial}{\partial t}}V=L_{\frac{\partial}{\partial \varphi}}V =L_{\frac{\partial}{\partial t}}S=L_{\frac{\partial}{\partial \varphi}}S=0$, getting restrictions on the spinor itself. In the present paper we focus on flagpole fields, whose form (up to the reversal of the third axis) is
\begin{eqnarray}\label{HPL2}
\psi=\left(\begin{tabular}{c}
$0$\\ $\Lambda\/e^{i\alpha}$\\ $\Xi\/e^{i\beta}$\\ $0$
\end{tabular}\right)
\end{eqnarray}
where $\Xi$ and $\Lambda$ are functions of the only variables $\rho$ and $z$, while $\alpha$ and $\beta$ are functions of all variables $\rho,z,\varphi,t$.

We stress that the form of the metric \eqref{Lewis} is compatible with the spinor \eqref{HPL2}, but more general axially-symmetric metrics are possible if the spinor is correspondingly enlarged. We are already approaching the problem of studying more general axially-symmetric exact solutions and we plan to write a further work on that. For the moment however, flagpole spinors are our interest, and so the metric \eqref{Lewis} is sufficiently general.

Evaluating the Dirac equations for the metric \eqref{Lewis} and for the spinor \eqref{HPL2}, and decomposing them in their real and imaginary parts, we get eight real equations of the form
\begin{subequations}\label{eqdd10ac}
\begin{equation}\label{eqdd10ac_1}
\frac{\Xi\cos\beta\left(PW\beta_\varphi + P\beta_t + \beta_\varphi\right)}{\sqrt{\rho\/P}}=0
\end{equation}
\begin{equation}\label{eqdd10ac_2}
\frac{\Xi\sin\beta\left(PW\beta_\varphi + P\beta_t + \beta_\varphi\right)}{\sqrt{\rho\/P}}=0
\end{equation}
\begin{equation}\label{eqdd10ac_3}
\begin{split}
\frac{1}{4\rho\/B^2}\left(-4\Xi\beta_\rho\/B\rho\cos\beta + 4\Xi\beta_z\/B\rho\sin\beta + \left(-4\Xi_\rho\/\rho\/B -2\Xi\left(\frac{W_\rho\/\rho\/BP}{2} + B_\rho\/\rho + B\right)\right)\sin\beta \right.\\
\left.- 4\rho\left(\left(\Xi_z\/B + \frac{\Xi\left(\frac{W_z\/BP}{2} + B_z\right)}{2}\right)\cos\beta + m\cos\alpha\/B^2\Lambda\right)\right)=0
\end{split}
\end{equation}
\begin{equation}\label{eqdd10ac_4}
\begin{split}
\frac{1}{4\rho\/B^2}\left(-4\Xi\beta_\rho\/B\rho\sin\beta - 4\Xi\beta_z\/B\rho\cos\beta + \left(4\Xi_\rho\/\rho\/B +2\Xi\left(\frac{W_\rho\/\rho\/BP}{2} + B_\rho\/\rho + B\right)\right)\cos\beta \right.\\
\left.- 4\rho\left(\left(\Xi_z\/B + \frac{\Xi\left(\frac{W_z\/BP}{2} + B_z\right)}{2}\right)\sin\beta + m\sin\alpha\/B^2\Lambda\right)\right)=0
\end{split}
\end{equation}
\begin{equation}\label{eqdd10ac_5}
\begin{split}
\frac{1}{4\rho\/B^2}\left(4\Lambda\alpha_\rho\/B\rho\cos\alpha + 4\Lambda\alpha_z\/B\rho\sin\alpha + \left(4\Lambda_\rho\/\rho\/B +2\Lambda\left(\frac{W_\rho\/\rho\/BP}{2} + B_\rho\/\rho + B\right)\right)\sin\alpha \right.\\
\left.- 4\rho\left(\left(\Lambda_z\/B + \frac{\Lambda\left(\frac{W_z\/BP}{2} + B_z\right)}{2}\right)\cos\alpha + m\cos\beta\/B^2\Xi\right)\right)=0
\end{split}
\end{equation}
\begin{equation}\label{eqdd10ac_6}
\begin{split}
\frac{1}{4\rho\/B^2}\left(4\Lambda\alpha_\rho\/B\rho\sin\alpha - 4\Lambda\alpha_z\/B\rho\cos\alpha + \left(-4\Lambda_\rho\/\rho\/B -2\Lambda\left(\frac{W_\rho\/\rho\/BP}{2} + B_\rho\/\rho + B\right)\right)\cos\alpha \right.\\
\left.- 4\rho\left(\left(\Lambda_z\/B + \frac{\Lambda\left(\frac{W_z\/BP}{2} + B_z\right)}{2}\right)\sin\alpha + m\sin\beta\/B^2\Xi\right)\right)=0
\end{split}
\end{equation}
\begin{equation}\label{eqdd10ac_7}
\frac{\Lambda\cos\alpha\left(PW\alpha_\varphi + P\alpha_t + \alpha_\varphi\right)}{\sqrt{\rho\/P}}=0
\end{equation}
\begin{equation}\label{eqdd10ac_8}
\frac{\Lambda\sin\alpha\left(PW\alpha_\varphi + P\alpha_t + \alpha_\varphi\right)}{\sqrt{\rho\/P}}=0
\end{equation}
\end{subequations}
and similarly we can evaluate the Einstein equations, which turn out to be
\begin{subequations}\label{eqcc10ac}
\begin{equation}\label{eqcc10ac_1}
\frac{-2P^2\/B_{\rho\rho}B\rho^2 - 2P^2\/B_{zz}B\rho^2 + 2P^2\/(B_\rho)^2\rho^2 + 2P^2\/BB_\rho\/\rho + 2 P^2\/(B_z)^2\rho^2 + B^2\left(\rho^2\/P^4\/(W_\rho)^2 - \rho^2\/(P_\rho)^2 + P^2\right)}{2P^2\rho^2\/B^2}
\end{equation}
\begin{equation}\label{eqcc10ac_2}
\frac{2P^2\/B_z + \rho\/B\left(W_\rho\/W_z\/P^4 - P_\rho\/P_z\right)}{2\rho\/BP^2} =0
\end{equation}
\begin{equation}\label{eqcc10ac_3}
\frac{1}{8\sqrt{P}B}\left(\sqrt{\rho}\left(2\Lambda^2\alpha_\rho\/BP + 2\Xi^2\beta_\rho\/BP + \left(PB_z - \frac{B\left(-P^2W_z + P_z\right)}{2}\right)\left(\Xi+\Lambda\right)\left(\Xi-\lambda\right)\right)\right)=0
\end{equation}
\begin{equation}\label{eqcc10ac_4}
\begin{split}
\frac{1}{8P^{3/2}B}\left(\sqrt{\rho}\left(2BP\Lambda^2\left(WP+1\right)\alpha_\rho + 2BP\Xi^2\left(WP+1\right)\beta_\rho + \left(\Xi+\Lambda\right)\left(\Xi-\Lambda\right)\left(\left(WP^2+P\right)B_z \right.\right.\right.\\
\left.\left.\left. - \frac{B\left(WP-1\right)\left(-P^2W_z + P_z\right)}{2}\right)\right)\right)=0
\end{split}
\end{equation}
\begin{equation}\label{eqcc10ac_5}
\frac{1}{2P^2B^2\rho}\left(-2P^2B_{\rho\rho}B\rho - 2P^2B_{zz}B\rho + 2P^2(B_\rho)^2\rho - 2P^2BB_\rho + \rho\left(B^2P^4(W_z)^2 - B^2(P_z)^2 + 2P^2(B_z)^2\right)\right)=0
\end{equation}
\begin{equation}\label{eqcc10ac_6}
\frac{1}{8\sqrt{P\rho}B}\left(-2\Lambda^2\alpha_zBP\rho - 2\Xi^2\beta_zBP\rho + \left(\Xi+\Lambda\right)\left(PB_\rho\rho - \frac{B\left(-P^2W_\rho\rho + P_\rho\rho + P\right)}{2}\right)\left(\Xi-\Lambda\right)\right)=0
\end{equation}
\begin{equation}\label{eqcc10ac_7}
\begin{split}
\frac{1}{8\sqrt{\rho}P^{3/2}B}\left(-2\rho\/BP\Lambda^2\left(WP+1\right)\alpha_z -2\rho\/BP\Xi^2\left(WP+1\right)\beta_z + \left(P\rho\left(WP+1\right)B_\rho \right.\right.\\
\left.\left. - \frac{B\left(\rho\left(WP-1\right)P_\rho - P\left(\rho\/P\left(WP-1\right)W_\rho - WP -1\right)\right)}{2}\right)\left(\Xi+\Lambda\right)\left(\Xi-\Lambda\right)\right)=0
\end{split}
\end{equation}
\begin{equation}\label{eqcc10ac_8}
\frac{1}{2B^2P}\left(-B^2\sqrt{\rho}\left(\Xi^2\beta_\varphi + \Lambda^2\alpha_\varphi\right)P^{3/2} - PP_{\rho\rho}\rho - PP_{zz}\rho + (P_\rho)^2\rho - PP_\rho + \left((P_z)^2 - P^4\left((W_\rho)^2 + (W_z)^2\right)\right)\rho\right)=0
\end{equation}
\begin{equation}\label{eqcc10ac_9}
\begin{split}
\frac{1}{4B^2P}\left(B^2\sqrt{\rho}\left(\Xi^2\beta_\varphi\/W + W\Lambda^2\alpha_\varphi - \Xi^2\beta_t - \Lambda^2\alpha_t\right)P^{3/2} + \alpha_\varphi\/B^2\Lambda^2\sqrt{\rho\/P} + \beta_\varphi\/B^2\Xi^2\sqrt{\rho\/P} + 2PWP_{\rho\rho}\rho \right.\\
+ 2PWP_{zz}\rho + 2P^2W_{\rho\rho}\rho + 2P^2W_{zz}\rho - 2W(P_\rho)^2\rho + 4P\left(\rho\/W_\rho + \frac{W}{2}\right)P_\rho + 2W(W_\rho)^2\rho\/P^4 + 2P^2W_\rho \\
\left. - 2\rho\left(-P^4W(W_z)^2 + W(P_z)^2 - 2PP_zW_z\right)\right)=0
\end{split}
\end{equation}
\begin{equation}\label{eqcc10ac_10}
\begin{split}
\frac{1}{2P^3B^2}\left(B^2\sqrt{\rho}\left(\Xi^2\beta_t + \Lambda^2\alpha_t\right)P^{5/2} + \sqrt{\rho}B^2W\left(\Xi^2\beta_t + \Lambda\alpha_t\right)P^{7/2} - \rho\/P\left(W^2P^2+1\right)P_{\rho\rho} - \rho\/P\left(W^2P^2+1\right)P_{zz}\right.\\
\left. - 2WW_{\rho\rho}\rho\/P^4 - 2WW_{zz}\rho\/P^4 + \rho\left(W^2P^2+1\right)(P_\rho)^2 + \left(-4\rho\/W_\rho\/P^3W - P^3W^2 -P\right)P_\rho - \rho\/P^4\left(W^2P^2+1\right)(W_\rho)^2 \right.\\
\left. - 2P^4WW_\rho + \left(\left(W^2P^2+1\right)(P_z)^2 - 4P_zW_zP^3W - (W_z)^2P^4\left(W^2P^2+1\right)\right)\rho\right)=0
\end{split}
\end{equation}
\end{subequations}
again as it is easy, although tedious, to check.

We are thus ready to proceed to the solution.
\section{Solving field equations}
In this section, we search for exact solutions of the field equations \eqref{eqdd10ac} and \eqref{eqcc10ac}.

To this end, we begin by deducing from the pairs of equations \eqref{eqdd10ac_1}--\eqref{eqdd10ac_2} and \eqref{eqdd10ac_7}--\eqref{eqdd10ac_8} the relations
\begin{equation}\label{sb34k}
\beta_t = - \frac{\beta_\varphi\left(WP+1\right)}{P}
\end{equation}
and 
\begin{equation}\label{sb24k}
\alpha_t = - \frac{\alpha_\varphi\left(WP+1\right)}{P}.
\end{equation}
After inserting equations \eqref{sb34k} and \eqref{sb24k} into equations \eqref{eqcc10ac}, from the linear combination $-W\eqref{eqcc10ac_8} - \eqref{eqcc10ac_9}$ we get
\begin{equation}\label{fW11kk}
W_{\rho\rho} = \frac{1}{\rho\/P^2}\left(-\Lambda^2\alpha_\varphi\/B^2\sqrt{\rho\/P} - \Xi^2\beta_\varphi\/B^2\sqrt{\rho\/P} - 2P_\rho\/W_\rho\/P\rho - 2P_z\/W_z\/P\rho - W_{zz}P^2\rho - W_\rho\/P^2\right)
\end{equation}
and using equation \eqref{fW11kk}, from equations \eqref{eqcc10ac_3} and \eqref{eqcc10ac_6} we obtain the identities
\begin{equation}\label{sb21k}
\alpha_\rho = - \frac{1}{4\Lambda^2BP}\left(-\Lambda^2BP^2W_z + \Xi^2BP^2W_z + 4\Xi^2\beta_\rho\/BP + \Lambda^2BP_z - 2\Lambda^2B_zP - \Xi^2BP_z + 2\Xi^2B_zP\right)
\end{equation}
and
\begin{equation}\label{sb22k}
\begin{split}
\alpha_z = \frac{1}{4\Lambda^2BP\rho}\left(-\Lambda^2BW_\rho\/P^2\rho + \Xi^2BW_\rho\/P^2\rho - 4\rho\/P\Xi^2\beta_zB + \Lambda^2BP_\rho\rho- 2\Lambda^2B_\rho\/P\rho - \Xi^2BP_\rho\rho \right.\\
\left. + 2\Xi^2B_\rho\/P\rho + \Lambda^2BP - \Xi^2BP\right).
\end{split}
\end{equation}
Inserting equations \eqref{sb21k} and \eqref{sb22k} as well as all the relations obtained above into the field equations, from the linear combination $\eqref{eqcc10ac_1}-\eqref{eqcc10ac_5}$ and from \eqref{eqcc10ac_2} we get the relations
\begin{equation}\label{vB1A}
B_\rho = \frac{B\left(\rho^2(W_z)^2P^4 - \rho^2(W_\rho)^2P^4 + \rho^2(P_\rho)^2 - \rho^2(P_z)^2 -P^2\right)}{4\rho\/P^2}
\end{equation}
and 
\begin{equation}\label{vB2A}
B_z = \frac{\rho\/B\left(-W_\rho\/W_zP^4 + P_\rho\/P_z\right)}{2P^2}.
\end{equation}
Collecting all the obtained results and inserting them into Dirac equations \eqref{eqdd10ac}, it is seen that the first two and last two of the latter are automatically satisfied while the remaining four can be expressed in the form
\begin{subequations}\label{eqdd10fc}
\begin{equation}\label{ssin23}
\begin{split}
\sin(\alpha-\beta) = \frac{1}{8\Lambda\/BP^2m\rho}\left(-\Xi(W_\rho)^2\rho^2P^4 + \Xi(W_z)^2\rho^2P^4 + 2\Xi\rho\/P^3W_\rho + \Xi\rho^2(P_\rho)^2 - \Xi\rho^2(P_z)^2 - 8\Xi\/P^2\beta_z\rho \right. \\
\left. + 8P^2\Xi_\rho\rho + 3P^2\Xi\right)
\end{split}
\end{equation}
\begin{equation}\label{scos23}
\cos(\alpha-\beta) = - \frac{-\Xi\/W_\rho\/W_z\rho\/P^4 + \Xi\/W_zP^3 + 4\Xi\beta_\rho\/P^2 + \Xi\/P_\rho\/P_z\rho + 4\Xi_zP^2}{4\Lambda\/BP^2m}
\end{equation}
\begin{equation}\label{ssin23a}
\begin{split}
\sin(\alpha-\beta) = \frac{1}{8\Xi\/mP^2B\Lambda\rho}\left( -\Xi^2(W_\rho)^2P^4\rho^2 + \Xi^2P^4(W_z)^2\rho^2 + 2\Xi^2W_\rho\/P^3\rho + \Xi^2(P_\rho)^2\rho^2 - \Xi^2(P_z)^2\rho^2 \right. \\
\left. - 8\rho\/P^2\Xi^2\beta_z + 2\Xi^2P_\rho\rho\/P + 8\Lambda\/P^2\Lambda_\rho\rho - 2\Xi^2P_\rho\rho\/P + 6\Lambda^2P^2 - 3\Xi^2P^2\right)
\end{split}
\end{equation}
\begin{equation}\label{scos23a}
\cos(\alpha-\beta) = - \frac{-\Xi^2W_\rho\/P^4W_z\rho + \Xi^2P^3W_z + 4\Xi^2\beta_\rho\/P^2 + \Xi^2P_\rho\/P_z\rho + \Lambda^2P_zP + 4\Lambda\/P^2\Lambda_z - \Xi^2P_zP}{4\Lambda\Xi\/BP^2m}.
\end{equation}
\end{subequations}
From the linear combinations $\eqref{scos23a}-\eqref{scos23}$ and $\eqref{ssin23a}-\eqref{ssin23}$ we get the relations
\begin{equation}\label{vdXi22}
\Lambda_z = - \frac{\Lambda^2P_z - \Xi^2P_z - 4\Xi\Xi_zP}{4\Lambda\/P}
\end{equation}
\begin{equation}\label{vdXi21}
\Lambda_\rho = - \frac{\Lambda^2P_\rho\rho - \Xi^2P_\rho\rho - 4\Xi\/P\Xi_\rho\rho + 3\Lambda^2P - 3\Xi^2P}{4\Lambda\/P\rho}.
\end{equation}
Without loss of generality, we can suppose that $\Lambda \geq \Xi$ and thus we can set
\begin{equation}\label{HH}
\Lambda = \sqrt{\Xi^2 + q^2e^{H(\rho,z)}}
\end{equation}
with $q$ a suitable constant. Substituting equation \eqref{HH} into \eqref{vdXi22} and \eqref{vdXi21}, we obtain the equations
\begin{subequations}\label{eqH}
\begin{equation}
H_\rho = - \frac{P_\rho\rho + 3P}{2\rho\/P}
\end{equation}
\begin{equation}
H_z = - \frac{P_z}{2P}
\end{equation}
\end{subequations}
which admit the solution
\begin{equation}\label{solH}
H(\rho,z) = - \frac{\ln\/P}{2} - \frac{3\ln\rho}{2} + \ln\/s
\end{equation}
for a suitable constant $s>0$. Inserting equation \eqref{solH} into equation \eqref{HH}, we have
\begin{equation}\label{solXi2}
\Lambda = \frac{1}{\rho^{3/4}}\sqrt{\frac{\Xi^2\sqrt{P}\rho^{3/2} + q^2s}{\sqrt{P}}}
\end{equation}
and using \eqref{solXi2} as well as of all the previously obtained identities, we can rewrite equation \eqref{eqcc10ac_8} in the form
\begin{equation}\label{eqfc8}
- \frac{\sqrt{\rho\/P}\alpha_\varphi\Xi^2}{2} - \frac{\sqrt{\rho\/P}\beta_\varphi\Xi^2}{2} - \frac{\rho\/P^3(W_\rho)^2}{2B^2} - \frac{\rho\/P^3(W_z)^2}{2B^2} - \frac{\alpha_\varphi\/q^2s}{2\rho} - \frac{\rho\/P_{\rho\rho}}{2B^2} - \frac{\rho\/P_{zz}}{2B^2} + \frac{\rho\/(P_\rho)^2}{2PB^2} + \frac{\rho(P_z)^2}{2PB^2} - \frac{P_\rho}{2B^2} =0.
\end{equation}
Differentiating equation \eqref{eqfc8} with respect to the variables $\varphi$ and $t$ and solving the consequent equations, we deduce the expression
\begin{equation}\label{rb23b}
\alpha = - \frac{\Xi^2\sqrt{P}\rho^{3/2}\beta}{\Xi^2\sqrt{P}\rho^{3/2} + q^2s} + Q_1(\rho,z)\varphi + Q_2(\rho,z,t)
\end{equation}
where (inserting equation \eqref{rb23b} into \eqref{eqfc8})
\begin{equation}\label{solQ1}
Q_1(\rho,z) = \frac{\rho\left(-PP_{\rho\rho}\rho - PP_{zz}\rho + (P_\rho)^2\rho - PP_\rho + \rho\left((P_z)^2 - P^4\left((W_\rho)^2 + (W_z)^2\right)\right)\right)}{B^2P\left(\Xi^2\sqrt{P}\rho^{3/2} + q^2s\right)}.
\end{equation}
Moreover, inserting equation \eqref{solQ1} into \eqref{sb24k} and making use of equation \eqref{sb34k}, we also get
\begin{equation}\label{solQ2k}
Q_2(\rho,z,t) = - \frac{(PW+1)Q_1(\rho,z)t}{P} + F(\rho,z)
\end{equation}
for an arbitrary function $F(\rho,z)$. Now, inserting equations \eqref{vB1A}, \eqref{vB2A}, \eqref{solXi2} and \eqref{rb23b} into \eqref{sb21k} and \eqref{sb22k}, and differentiating the resulting equations twice with respect to the variables $t$ and $\varphi$, we obtain the equations	
\begin{subequations}\label{6.1.62}
\begin{equation}
\frac{sq^2\Xi\sqrt{\rho}\beta_{tt}\left(4P\rho\Xi_\rho + P_\rho\Xi\rho + 3P\Xi\right)}{2\sqrt{P}\left(\Xi^2\sqrt{P}\rho^{3/2} + q^2s\right)^2} =0
\end{equation}
\begin{equation}
\frac{\Xi\rho^{3/2}\beta_{tt}q^2s\left(4P\Xi_z + \Xi\/P_z\right)}{2\sqrt{P}\left(\Xi^2\sqrt{P}\rho^{3/2} + q^2s\right)^2} =0
\end{equation}
\end{subequations}
\begin{subequations}\label{6.1.63}
\begin{equation}
\frac{sq^2\Xi\sqrt{\rho}\beta_{\varphi\varphi}\left(4P\rho\Xi_\rho + P_\rho\Xi\rho + 3P\Xi\right)}{2\sqrt{P}\left(\Xi^2\sqrt{P}\rho^{3/2} + q^2s\right)^2} =0
\end{equation}
\begin{equation}
\frac{\Xi\rho^{3/2}\beta_{\varphi\varphi}q^2s\left(4P\Xi_z + \Xi\/P_z\right)}{2\sqrt{P}\left(\Xi^2\sqrt{P}\rho^{3/2} + q^2s\right)^2} =0
\end{equation}
\end{subequations}
\begin{subequations}\label{6.1.64}
\begin{equation}
\frac{sq^2\Xi\sqrt{\rho}\beta_{t\varphi}\left(4P\rho\Xi_\rho + P_\rho\Xi\rho + 3P\Xi\right)}{2\sqrt{P}\left(\Xi^2\sqrt{P}\rho^{3/2} + q^2s\right)^2} =0
\end{equation}
\begin{equation}
\frac{\Xi\rho^{3/2}\beta_{t\varphi}q^2s\left(4P\Xi_z + \Xi\/P_z\right)}{2\sqrt{P}\left(\Xi^2\sqrt{P}\rho^{3/2} + q^2s\right)^2} =0.
\end{equation}
\end{subequations}
The previous equations \eqref{6.1.62}, \eqref{6.1.63} and \eqref{6.1.64} highlight the following three main cases:
\begin{enumerate}
	\item $q=0 \quad \Longleftrightarrow \quad \Lambda=\Xi$ 
	\item $\beta_{tt}=0, \quad \beta_{t\varphi}=0, \quad \beta_{\varphi\varphi}=0$
	\item $4P\rho\Xi_\rho + P_\rho\Xi\rho + 3P\Xi =0, \quad 4P\Xi_z + \Xi\/P_z =0$ 
\end{enumerate}
together with all their possible intersections. Omitting the details for the sake of brevity, we have that cases $2$ and $3$ with $\Lambda \not = \Xi$ have no solutions. We restrict our attention to the intersection between case $1$ and $2$, postponing the study of the general case $1$ to a future work. Therefore, from now on we simultaneously assume the conditions 
\begin{equation}\label{case12}
\Lambda=\Xi, \qquad \beta_{tt}=0, \qquad \beta_{t\varphi}=0, \qquad \beta_{\varphi\varphi}=0
\end{equation}
and then, taking conditions \eqref{case12} into account, from equation \eqref{sb34k} we have
\begin{equation}\label{sb3c20}
\beta = M(\rho,z)t - \frac{M(\rho,z)P(\rho,z)\varphi + \phi(\rho,z)}{PW+1}
\end{equation}
where $M(\rho,z)$ and $\phi(\rho,z)$ are two arbitrary functions of the variables $\rho$ and $z$. At the same time, differentiating equations \eqref{sb21k} and \eqref{sb22k} with respect to $\varphi$ after inserting all the stated results, we obtain the relations
\begin{equation}\label{sQ120}
(Q_1)_\rho =0, \quad (Q_1)_z =0 \quad \Rightarrow \quad Q_1(\rho,z)=k
\end{equation}
where $k$ is a constant. Repeating the procedure, but this time differentiating equations \eqref{sb21k} and \eqref{sb22k} with respect to the variable $t$ and using equation \eqref{sQ120}, we get also
\begin{equation}\label{solWcp20}
W = -\frac{1}{P} + c
\end{equation}
with $c$ constant. Inserting equations \eqref{sQ120} and \eqref{solWcp20} into \eqref{sb21k} and \eqref{sb22k}, we derive the relations
\begin{equation}\label{sFac2}
F_\rho =0, \quad F_z =0 \quad \Rightarrow \quad F(\rho,z)=h
\end{equation}
with $h$ constant. Using equations \eqref{solXi2}, \eqref{rb23b}, \eqref{solQ2k}, \eqref{sb3c20} and \eqref{solWcp20}, we can solve equation \eqref{eqcc10ac_2} and the linear combination $\eqref{eqcc10ac_1} - \eqref{eqcc10ac_5}$ for the unknown $B(\rho,z)$ and obtain the expression
\begin{equation}\label{solBc20}
B = \frac{b}{\rho^{1/4}}
\end{equation} 
with $b$ constant. Replacing equations \eqref{sQ120}, \eqref{solWcp20} and \eqref{solBc20} into equation \eqref{solQ1}, we obtain the following differential equation
\begin{equation}\label{sP11c20}
P_{\rho\rho} + \frac{1}{\rho}P_\rho + P_{zz} = -\frac{k\sqrt{P}b^2\Lambda^2}{\rho}
\end{equation} 
identical to the Poisson equation for the unknown $P(\rho,z)$ in cylindrical coordinates. Moreover, from equations \eqref{rb23b} and \eqref{sb3c20} together with equations \eqref{solQ2k}, \eqref{sQ120}, \eqref{solWcp20}, \eqref{sFac2} we get the expressions
\begin{equation}\label{sb2c20}
\alpha = -ckt +k\varphi - \beta +h
\end{equation}
and
\begin{equation}\label{sb3c200}
\beta = \frac{\left(ct-\varphi\right)M(\rho,z) + c\phi(\rho,z)}{c}.
\end{equation}
Collecting all the results obtained so far, we have the Lewis-Papapetrou metric \eqref{Lewis} and the spinor field \eqref{HPL2} respectively expressed as 
\begin{equation}\label{Lewis2}
{ds}^{2}= -\frac{b^2(d\rho^2 +dz^2)}{\sqrt{\rho}} -\rho\/P \left(-\left(-\frac{1}{P} +c\right)\,dt + d\varphi\right)^2 + \frac{\rho}{P}\,dt^2
\end{equation}
and
\begin{eqnarray}\label{HPL22}
\psi=\left(\begin{tabular}{c}
$0$\\ $\Lambda\/e^{i( -ckt +k\varphi - \beta +h)}$\\ $\Lambda\/e^{i\frac{\left(ct-\varphi\right)M(\rho,z) + c\phi(\rho,z)}{c}}$\\ $0$
\end{tabular}\right)
\end{eqnarray}
and as a direct check would show, the metric tensors and spinor fields of the form \eqref{Lewis2} and \eqref{HPL22} satisfy all the Einstein--like field equations \eqref{eqcc10ac}. In particular, the function $P(\rho,z)$ appearing in equation \eqref{Lewis2} has to satisfy the Poisson equation \eqref{sP11c20}, while in order to determine the remaining functions $\Lambda(\rho,z)$, $M(\rho,z)$ and $\phi(\rho,z)$ in equation \eqref{HPL22} we still have at our disposal four of the eight Dirac equations \eqref{eqdd10ac}. About this, it is convenient assuming the function $\Lambda(\rho,z)$ of the form
\begin{equation}\label{ndx3}
\Lambda = \frac{\Psi}{P^{\frac{1}{4}}\rho^{3/4}}
\end{equation}
where $\Psi(\rho,z)$ is a function of the variables $\rho$ and $z$. After that, evaluating equations \eqref{eqdd10ac_3} and \eqref{eqdd10ac_4} for the metric \eqref{Lewis2} and the spinor field \eqref{HPL22}, we get the equations
\begin{subequations}\label{fsincosz0}
\begin{equation}\label{fsincosz0_1}
\sin(\beta + (ct-\varphi)k -h)= \frac{\left(\rho\Psi\beta_\rho\sin\beta + \rho\Psi\beta_z\cos\beta + \left(-\Psi_\rho\rho + \frac{3\Psi}{8}\right)\cos\beta + \rho\Psi_z\sin\beta\right)}{\rho^{3/4}bm\Psi}
\end{equation}
\begin{equation}\label{fsincosz0_2}
\cos(\beta + (ct-\varphi)k -h)= - \frac{\left(\rho\Psi\beta_\rho\cos\beta - \rho\Psi\beta_z\sin\beta + \left(\Psi_\rho\rho - \frac{3\Psi}{8}\right)\sin\beta + \rho\Psi_z\cos\beta\right)}{\rho^{3/4}bm\Psi}.
\end{equation}
\end{subequations}
From linear combinations of \eqref{fsincosz0} we deduce the equations
\begin{subequations}\label{eqddw0}
\begin{equation}\label{eqddw0_1}
\sin(2\beta + (ct-\varphi)k -h)= \frac{8\Psi\beta_z\rho - 8\Psi_\rho\rho + 3\Psi}{8\Psi\rho^{3/4}mb}
\end{equation}
\begin{equation}\label{eqddw0_2}
\cos(2\beta + (ct-\varphi)k -h)= - \frac{\left(\Psi\beta_\rho + \Psi_z\right)\rho^{1/4}}{\Psi\/mb}
\end{equation}
\end{subequations}
and by multiplying equation \eqref{eqddw0_1} for $(2\beta_\varphi -k)$ and adding the result to the derivative of equation \eqref{eqddw0_2} with respect to $\varphi$, we obtain the equation
\begin{equation}\label{sb13c20}
\beta_{\rho\varphi} = - \frac{\left(\Psi\beta_z\rho - \Psi_\rho\rho + \frac{3\Psi}{8}\right)\left(k - 2\beta_\varphi\right)}{\rho\Psi}.
\end{equation}
Analogously, by subtracting the derivative of equation \eqref{eqddw0_1} with respect to $\varphi$ to equation \eqref{eqddw0_2} multiplied for $(2\beta_\varphi -k)$, we also obtain 
\begin{equation}\label{sb23c20}
\beta_{z\varphi} = \frac{\left(\Psi\beta_\rho + \Psi_z\right)\left(k - 2\beta_\varphi\right)}{\Psi}.
\end{equation}
Inserting expression \eqref{sb3c200} into equations \eqref{sb13c20} and \eqref{sb23c20} and differentiating with respect to $\varphi$, we get the relations
\begin{subequations}\label{M1M2}
\begin{equation}\label{M1}
\frac{M_z\left(k + \frac{2M}{c}\right)}{c} =0
\end{equation}
\begin{equation}\label{M2}
\frac{M_\rho\left(k + \frac{2M}{c}\right)}{c} =0
\end{equation}
\end{subequations}
so that from equations \eqref{M1M2}, we get the identity
\begin{equation}\label{vkc20}
M(\rho,z)= M := -\frac{ck}{2}.
\end{equation}
In view of equation \eqref{vkc20}, the expressions \eqref{sb2c20} and \eqref{sb3c200} for the functions $\alpha$ and $\beta$ assume the form
\begin{equation}\label{sb2cp20a}
\alpha = Mt - \frac{M\varphi}{c} - \phi(\rho,z) + h
\end{equation}
\begin{equation}\label{sb3cp20a}
\beta = Mt- \frac{M\varphi}{c} + \phi(\rho,z).
\end{equation}
Inserting expressions \eqref{sb3cp20a} into equations \eqref{eqddw0} and resolving the resultant equations for $\Psi_\rho$ and $\Psi_z$, we obtain the relations
\begin{subequations}\label{fdX3122a}
\begin{equation}\label{fdX3122a_1}
\Psi_\rho = \frac{\Psi\left(8\sin(-2\phi +h)\rho^{3/4}bm + 8\phi_z\rho +3\right)}{8\rho}
\end{equation}
\begin{equation}\label{fdX3122a_2}
\Psi_z = - \frac{\Psi\left(\cos(-2\phi + h)bm + \phi_\rho\rho^{1/4}\right)}{\rho^{1/4}}.
\end{equation}
\end{subequations}
From equations \eqref{fdX3122a}, by imposing Schwartz conditions, we get the following differential equation for the function $\phi$
\begin{equation}\label{eqphic20}
\phi_{\rho\rho} + \phi_{zz} + \frac{\left(-2mb\left(\phi_z\rho + \frac{1}{8}\right)\cos(-2\phi + h) + 2\phi_\rho\sin(-2\phi + h)bm\rho\right)}{\rho^{5/4}} = 0
\end{equation}
which can be rewritten in the form
\begin{equation}\label{eqphic20mm}
- \frac{\partial}{\partial\rho}\left(\phi_\rho + \frac{\cos(-2\phi + h)bm}{\rho^{1/4}}\right) =
\frac{\partial}{\partial z}\left(\phi_z + \frac{\sin(-2\phi + h)bm}{\rho^{1/4}}\right).
\end{equation}
From equation \eqref{eqphic20mm}, it follows that there exist a function $T(\rho,z)$ such that
\begin{subequations}\label{fdTau12}
\begin{equation}\label{fdTau12_1}
T_\rho = \phi_z + \frac{\sin(-2\phi + h)bm}{\rho^{1/4}}
\end{equation}
\begin{equation}\label{fdTau12_2}
T_z = - \phi_\rho - \frac{\cos(-2\phi + h)bm}{\rho^{1/4}}.
\end{equation}
\end{subequations}
Substituting the content of equations \eqref{fdTau12} into equations \eqref{fdX3122a}, we deduce the equations
\begin{subequations}\label{sisLa}
\begin{equation}\label{sisLa_1}
\Psi_\rho = \frac{\left(8T_\rho\rho + 3\right)\Psi}{8\rho}
\end{equation}
\begin{equation}\label{sisLa_2}
\Psi_z = \Psi\/T_z
\end{equation}
\end{subequations}
which can be solved for the function $\Psi$, giving rise to the expression
\begin{equation}\label{solLac2}
\Psi(\rho,z) = C_\Psi\rho^{3/8}e^{T(\rho,z)} 
\end{equation}
where $C_\Psi$ is a suitable constant. 

Summarizing all the obtained results, we have found solutions of the field equations \eqref{eqdd10ac} and \eqref{eqcc10ac} of the form
\begin{equation}\label{Lewis3}
{ds}^{2}= -\frac{b^2(d\rho^2 +dz^2)}{\sqrt{\rho}} -\rho\/P \left(-\left(-\frac{1}{P} +c\right)\,dt + d\varphi\right)^2 + \frac{\rho}{P}\,dt^2
\end{equation}
and
\begin{eqnarray}\label{HPL23}
\psi=\left(\begin{tabular}{c}
$0$\\ 
\\
$\frac{C_\Psi\/e^{\frac{i\left(Mct -M\varphi -c\phi(\rho,z) + ch\right)+ cT(\rho,z)}{c}}}{\rho^{3/8}P^{1/4}}$\\ 
\\
$\frac{C_\Psi\/e^{\frac{i\left(c\phi(\rho,z) + Mct -M\varphi\right)+ cT(\rho,z)}{c}}}{\rho^{3/8}P^{1/4}}$\\ 
\\
$0$ 
\end{tabular}\right)
\end{eqnarray}
where the functions $T(\rho,z)$ and $P(\rho,z)$ are determined in terms of the function $\phi(\rho,z)$, solution of equation \eqref{eqphic20}, by means of equations \eqref{fdTau12} and equation \eqref{sP11c20} now expressed as 
\begin{equation}\label{Poisson}
P_{\rho\rho} + \frac{1}{\rho}P_\rho + P_{zz} = \frac{2Mb^2C_\Psi^2e^{2T}}{c\rho^{7/4}}
\end{equation} 
The proposed resolution of the field equations \eqref{eqdd10ac} and \eqref{eqcc10ac} is therefore entirely based on particular solutions of equation \eqref{eqphic20}. In the next subsections we present two examples relative to two choices of the function $\phi$.

We notice that, regardless of the particular solutions of equation \eqref{eqphic20} and related equations \eqref{fdTau12} and \eqref{sP11c20}, the Riemann invariants associated with the solutions \eqref{Lewis3} and \eqref{HPL23} are of the form
\begin{equation}\label{Invariants}
S_1 = \frac{1}{64b^4\rho^3}, \quad S_2 = \frac{1}{512b^6\rho^{9/2}}
\end{equation}
thus presenting a singularity on the submanifold $\rho=0$.

As far as the spinor field is concerned, the velocity vector (tetrad) components result to be
\begin{equation}\label{momentum_components}
\bar\psi\gamma^1\psi = \bar\psi\gamma^2\psi =0, \quad \bar\psi\gamma^3\psi = \bar\psi\gamma^4\psi = \frac{2C_\Psi^2e^{2T}}{\rho^{3/4}\sqrt{P}}
\end{equation}
with Pauli-Lubanski axial-vector and both scalar bispinor quantities equal to zero, confirming that the spinor is a flagpole spinor, belonging to the type-IV class, as we have claimed from the start.
\subsection{The case \texorpdfstring{$\phi$}{constant} constant}
Supposing $\phi$ constant, from equations \eqref{eqphic20} and \eqref{fdTau12} we have the relations
\begin{equation}\label{sfa}
h= 2\phi -\frac{\pi}{2}
\end{equation} 
and
\begin{equation}\label{ex1T}
T(\rho) = - \frac{4\rho^{3/4}bm}{3}
\end{equation}
where the choice $2\phi -h= \pi/2$ is justified by the requirement of convergence, for large values of $\rho$, of the source term in the Poisson equation \eqref{Poisson}, now written in the form
\begin{equation}\label{Poissonex1}
P_{\rho\rho} + \frac{1}{\rho}P_\rho + P_{zz} = \frac{2Mb^2C_\Psi^2e^{- \frac{8\rho^{3/4}bm}{3}}}{c\rho^{7/4}}
\end{equation}
and the whole procedure would end by finding some solutions of equation \eqref{Poissonex1} and inserting them into expressions \eqref{Lewis3} and \eqref{HPL23}. To this end, for simplicity, let us first rename some constants as
\begin{equation}\label{vab}
C_1:=\frac{2Mb^2C^2_\Psi}{c} \quad {\rm and} \quad C_2:= \frac{8bm}{3}
\end{equation}
and we search for a particular radial solution $\tilde{P}(\rho)$ of equation \eqref{Poissonex1} by setting $p(\rho):= \tilde{P}_\rho$ and solving the resultant equation
\begin{equation}\label{eqpex}
p_\rho + \frac{p}{\rho} = \frac{C_1e^{-C_2\rho^{3/4}}}{\rho^{7/4}}.
\end{equation}
A solution of equation \eqref{eqpex} is given by
\begin{equation}\label{spex10b}
p(\rho) = \frac{8\pi\sqrt{3}C_1}{9\Gamma(2/3)C_2^{1/3}\rho} - \frac{4C_1\Gamma(1/3,C_2\rho^{3/4})}{3C_2^{1/3}\rho}
\end{equation}
where $\Gamma(-)$ and $\Gamma(-,-)$ are the Gamma function and the Incomplete Gamma function respectively \cite{Stegun}. It is seen that a primitive function of \eqref{spex10b} is given by
\begin{equation}\label{soPP}
\tilde{P}(\rho) = \frac{8\pi\sqrt{3}C_1\ln\rho}{9\Gamma(2/3)C_2^{1/3}} + 16C_1\rho^{1/4}\,_2\/F_2(1/3,1/3;4/3,4/3;-C_2\rho^{3/4})
\end{equation}
where $\,_2\/F_2(1/3,1/3;4/3,4/3;-)$ is a generalized hypergeometric function of kind $\,_2\/F_2$. The function \eqref{soPP} is then a particular solution of equation \eqref{Poissonex1}. We notice that the term proportional to $\ln\rho$ can be omitted since it is harmonic, thus ensuring the positiveness of the solution even for small values of $\rho$, as required by the expressions \eqref{Lewis3} and \eqref{HPL23}. Of course, we can generate other solutions by adding to the function $\tilde{P}$ some generic harmonic functions $\bar{P}(\rho,z)$ in the cylindrical coordinates $\rho$ and $z$, as for example
\begin{equation}\label{genericsolution}
P(\rho,z) = \tilde{P}(\rho) + \bar{P}(\rho,z)
\end{equation}
with $\bar{P}(\rho,z) = \frac{K}{\sqrt{\rho^2 + z^2}}$. The final expression of the spinor field is
\begin{eqnarray}\label{spinoreex1}
\psi=\left(\begin{tabular}{c}
$0$\\ 
\\
$\frac{C_\Psi\/e^{-\frac{C_2}{2}\rho^{3/4}}e^{i\left(Mt-M\varphi/c +\phi -\pi/2\right)}}{\rho^{3/8}P^{1/4}}$\\ 
\\
$\frac{C_\Psi\/e^{-\frac{C_2}{2}\rho^{3/4}}e^{i\left(Mt-M\varphi/c +\phi \right)}}{\rho^{3/8}P^{1/4}}$\\ 
\\
$0$ 
\end{tabular}\right).
\end{eqnarray}
The velocity vector (tetrad) components of the spinor \eqref{spinoreex1} are
\begin{equation}\label{spinorcomponentssex1}
\bar\psi\gamma^1\psi = \bar\psi\gamma^2\psi =0, \quad \bar\psi\gamma^3\psi= \bar\psi\gamma^4\psi = \frac{2C_\Psi^2e^{-C_2\rho^{3/4}}}{\rho^{3/4}\sqrt{P}}
\end{equation}
and the metric tensor is given by equation \eqref{Lewis2}.
\subsection{The case \texorpdfstring{$\phi$}{radial} radial}
We suppose that $\phi$ is function of the single radial variable $\rho$. In such a circumstance, setting $h=0$ for simplicity, equation \eqref{eqphic20} assumes the form
\begin{equation}\label{equazionephi}
\phi_{\rho\rho} = \frac{bm\left(8\rho\phi_\rho\sin(2\phi)+\cos(2\phi)\right)}{4\rho^{5/4}}
\end{equation}
with a solution of equation \eqref{equazionephi} given by 
\begin{equation}\label{ipp}
\phi(\rho) = \frac{1}{2}\arctan\left(\frac{e^{-2C_2\rho^{3/4}}-1}{2e^{-C_2\rho^{3/4}}}\right).
\end{equation}
Inserting the expression \eqref{ipp} into equation \eqref{fdTau12} and solving for $T(\rho)$, we get 
\begin{equation}\label{sTauco}
T(\rho) = - \frac{C_2\rho^{3/4}-1}{2} + \ln\/C_3
\end{equation}
where $C_3$ is a suitable integration constant. As done in the previous subsection, we set $p(\rho):=P_\rho(\rho)$ and rewrite equation \eqref{Poisson} as
\begin{equation}\label{eqp1}
p_\rho + \frac{p}{\rho} - \frac{2C_1C_3^2\cosh\left(C_2\rho^{3/4}\right)}{\rho^{7/4}}
\end{equation}
equation \eqref{eqp1} admits the solution
\begin{equation}\label{sopcp}
p(\rho) = - \frac{3^{1/3}4C_1C_3^2\Gamma\left(\frac{1}{3},-C_2\rho^{3/4}\right)}{3\left(1+i\sqrt{3}\right)b^{1/3}m^{1/3}\rho}
- \frac{2\left(3^{1/3}+i3^{5/6}\right)C_1C_3^2\Gamma\left(\frac{1}{3},C_2\rho^{3/4}\right)}{3\left(1+i\sqrt{3}\right)b^{1/3}m^{1/3}\rho} + \frac{C_4}{\rho}
\end{equation}
where again $\Gamma(-)$ and $\Gamma(-,-)$ denote the Gamma and the Incomplete Gamma function respectively, and $C_4$ is an integration constant. By integrating with respect to $\rho$ and omitting all the additive harmonic terms, we get a particular solution of the Poisson equation \eqref{Poisson}, expressed as
\begin{equation}\label{soPPgg}
\tilde{P}(\rho) = 16C_1C_3^2\rho^{1/4}\/_2F_2\left(\frac{1}{3},\frac{1}{3};\frac{4}{3},\frac{4}{3};-C_2\rho^{3/4}\right) +
16C_1C_3^2\rho^{1/4}\/_2F_2\left(\frac{1}{3},\frac{1}{3};\frac{4}{3},\frac{4}{3};C_2\rho^{3/4}\right)
\end{equation}
where, as above, $\,_2\/F_2(\frac{1}{3},\frac{1}{3};\frac{4}{3},\frac{4}{3};-)$ is a generalized hypergeometric function of kind $\,_2\/F_2$. Again, by adding to $\tilde{P}(\rho)$ any harmonic function $\bar{P}(\rho,z)$, we obtain solutions $P(\rho,z) = \tilde{P}(\rho) + \bar{P}(\rho,z)$ of equation \eqref{Poisson} in the variables $\rho$ and $z$. Of course, care must be taken that the so obtained solution $P(\rho,z)$ is positive (solution \eqref{soPPgg} actually is positive). The explicit form of the spinor field is given by
\begin{eqnarray}\label{spinoreex2}
\psi=\left(\begin{tabular}{c}
$0$\\ 
\\
$\frac{C_\Psi\sqrt{2\cosh\left(C_2\rho^{3/4}\right)}C_3e^{i\left(\frac{\left(ct-\varphi\right)M}{c}+\frac{\arctan\left(\sinh\left(C_2\rho^{3/4}\right)\right)}{2}\right)}}{\rho^{3/8}P^{1/4}}$\\ 
\\
$\frac{C_\Psi\sqrt{2\cosh\left(C_2\rho^{3/4}\right)}C_3e^{i\left(\frac{\left(ct-\varphi\right)M}{c}-\frac{\arctan\left(\sinh\left(C_2\rho^{3/4}\right)\right)}{2}\right)}}{\rho^{3/8}P^{1/4}}$\\ 
\\
$0$ 
\end{tabular}\right).
\end{eqnarray}
The velocity vector (tetrad) components of the spinor \eqref{spinoreex2} are
\begin{equation}\label{spinorcomponentssex2}
\bar\psi\gamma^1\psi = \bar\psi\gamma^2\psi =0, \quad \bar\psi\gamma^3\psi= \bar\psi\gamma^4\psi = \frac{4C_\Psi^2\/C_3^2\cosh\left(C_2\rho^{3/4}\right)}{\rho^{3/4}\sqrt{P}}
\end{equation}
with metric tensor given by equation \eqref{Lewis2}.
\section{Conclusion}
In this paper, we have presented an exact solution for an interacting system of a spinor in its own gravitational field in which the gravitational field was acting back onto the matter distribution itself. This fully-coupled system of self-gravitating matter was studied in the case of flagpole spinors. Two notable sub-cases were analyzed in detail.

Because flagpole fields have Pauli-Lubanski axial-vector equal to zero identically, it turns out that they can have no torsion either, at least in Einstein-Sciama-Kibble theories. Notice that even if we were to have a propagating torsion field, it would still lack an interaction to flagpole spinors. So, whether gravity is completed with torsion in a non-propagating or a propagating way, torsion may be zero or non-zero, but either way it will not couple to the spin of matter fields. Thus, there appears to be no possible generalization for the background (unless of course the generalization comes from the inclusion of an entirely new type of physical fields and their interactions).

Feasible extensions can however be obtained by enlarging the type of spinors to be flag-dipole spinors. This solution would constitute the most comprehensive type of spinors of singular type, and therefore it is worth to look for it.

The entire set of solutions that we have found so far in this paper and in \cite{CFV1} consists of flagpole and Weyl fields while those found in \cite{CFV2} are Dirac fields with the critical property of having scalar bilinear equal to zero: therefore, all solutions found so far share the property of having $\overline{\psi}\psi\!\equiv\!0$ in a way or another. Because what is usually done in quantum field theory consists in developing in plane waves, for which the pseudo-scalar bilinear may vanish but the scalar bilinear must be a non-zero constant, then we can conclude that all our solutions do not possess at least some of the basic properties that would allow them to be treated according to the usual methods employed in quantum field theory. Still, all of them are exact solutions. Does this mean that the system of field equations is not sufficient, taken alone, to single out all physical solutions, and some of them do exist which are nevertheless not physical? Or does it mean that the methods of QFT are too restrictive to deal with fields that would otherwise be correctly defined?

Continuing the research along this line is a way to furnish answers to questions that, involving both classical exact solutions and quantization methods, lie at the very center of the foundations of contemporary physics.

\end{document}